\documentclass{aa}  
\usepackage{graphicx}
\usepackage{txfonts}
\usepackage{natbib}
\usepackage[bookmarks = true,colorlinks = true,linkcolor = blue,urlcolor = blue,citecolor = blue,breaklinks]{hyperref}

\graphicspath{ {./figures/} }

\newcommand\ms[1]{m s$^{-1}$}

\newcommand{\kms}{km s$^{-1}$} 
\newcommand{\degree}{^\circ}

\newcommand{\eg}{e.g., }

\newcommand{\ie}{\textit{i.e.,}}

\newcommand{\ha}{H$\alpha$}

\begin{document} 

 \title{Observations of a prominence eruption and loop contraction}

\author{Pooja Devi
\inst{1}\and
Pascal Démoulin
\inst{2,3}\and
Ramesh Chandra 
\inst{1}\and
Reetika Joshi
\inst{1}\and
Brigitte Schmieder
\inst{2,4}\and
Bhuwan Joshi
\inst{5}
}
\institute
{Department of Physics, DSB Campus, Kumaun University, Nainital -- 263 001, India\\
\email{setiapooja.ps@gmail.com}
\and
LESIA, Observatoire de Paris, Universit\'e PSL, CNRS, Sorbonne Universit\'e, Universit\'e de Paris, 5 place Jules Janssen, F-92190 Meudon, France
\and
 Laboratoire Cogitamus, 1 3/4 rue Descartes, 75005 Paris, France
\and
Centre for Mathematical Plasma Astrophysics, Dept. of Mathematics, KU Leuven, 3001 Leuven, Belgium
 \and
Udaipur Solar Observatory, Physical Research Laboratory, Udaipur 313004, India
}

\authorrunning{Pooja Devi et al.} 
   \titlerunning{Prominence eruption and loop contraction}
\abstract
{Prominence eruptions provide key observations to understand the launch of coronal mass ejections as their cold plasma traces a part of the unstable magnetic configuration.}
{We select a well observed case to derive observational constraints for eruption models. 
}
{We analyze the prominence eruption and loop
expansion and contraction observed on 02 March 2015 associated with a GOES M3.7 class flare (SOL2015-03-02T15:27) using the data from Atmospheric Imaging Assembly (AIA) and the Reuven Ramaty High Energy Solar Spectroscopic Imager (RHESSI).  We study the prominence eruption and the
 evolution of loops using the time-distance techniques. 
}
 {  The source region is a decaying bipolar active region where magnetic flux cancellation is present for several days before the eruption. AIA observations locate the erupting prominence within a flux rope viewed along its local axis direction. We identify and quantify the motion of loops in contraction and expansion located on the side of the erupting flux rope. Finally, 
RHESSI hard X-ray observations identify the loop top and two foot-point sources.
}
{
Both AIA and RHESSI observations support the standard model of eruptive flares. The contraction occurs 19 minutes after the start of the prominence eruption indicating that this contraction is not associated with the eruption driver.  Rather, this prominence eruption is compatible with an unstable flux rope where the contraction and expansion of the lateral loop is the consequence of a side vortex developing after the flux rope is launched.
}

  \keywords{Sun: filaments, prominences -- Sun: magnetic fields -- Sun: flares
          }

   \maketitle
\section{Introduction}
\label{sect_Introduction}
Solar prominence, or filament, eruptions are one of the violent illustrations of solar activity. Because of their associations with 
coronal mass ejections (CMEs) and interplanetary coronal mass ejections (ICMEs), their study is important for the space-weather point of view 
\citep[for example see: ][]{ Schwenn2006, Verbanac2011, Schmieder2020}.
In the standard flare model, also called the CSHKP model \citep{Carmichael64, Sturrock66, Hirayama74, Kopp76}, the ejection of a flux rope (FR), possibly containing a filament, 
drives the magnetic reconnection behind it which implies a two-ribbon flare.
In the FR, the magnetic field lines have helical structures and the dense as well as cold prominence plasma material is concentrated in magnetic dips.
Therefore, the existence of prominence can be considered as an indicator of the magnetic
FR in solar corona \citep{Schmieder2013, Filippov2015}.
The relationship between the FR rise and associated radiative
signatures has been a topic of considerable interest
\citep[\eg ][]{Alexander2006, Liu2009bj, Joshi2013, Joshi2016, Mitra2019}.
In particular, comparisons of the location, timing, and strength of high-energy emissions
(e.g., temporal and spatial evolution of HXR sources) with respect to the
dynamical evolution of the prominence provide critical clues to help understand the
characteristics of the underlying energy release phenomena, such as the expected site of magnetic reconnection, particle acceleration, and heating.

Two- and three-dimensional models have been proposed to explain prominence and filament eruptions and their associated activities such as the spatial location and separation of flare ribbons.  
Models have also been proposed to explain the triggering mechanism of the eruptions, 
for example, the magnetic breakout \citep{Antiochos1998,Antiochos1999}, the tether cutting \citep{Moore2001, Moore2006}, the kink instability \citep{Sakurai1976,Torok2005}, and the torus instability \citep{Forbes1991, Kliem2006} models.
The instability of an FR, modeled by a line current in equilibrium in a bipolar potential field, was first proposed by \citet{VanTend1978} to model solar eruptions.
This model was further developed toward an MHD model \citep[\eg ][]{Demoulin1991, Forbes1991, Lin2000, Priest2002, Forbes2006, Kliem2006, Aulanier2010}.
It is presently known as the torus instability model, or equivalently the catastrophe model \citep{Demoulin2010}.
This catastrophic instability occurs in a bipolar magnetic field configuration with an FR in equilibrium above the photospheric inversion line (PIL) of the magnetic field vertical component. 
This model is tied to the decrease in the magnetic field strength with height that implies a decrease in the downward and stabilizing force as the FR is forced to evolve to a larger height.  
This evolution is typically generated by the photospheric evolution 
(\eg new magnetic flux emergence, shearing, and/or converging motions around the PIL). 
At some critical height, the FR become unstable and erupts. 
 
During the FR eruption, the arcade-like magnetic field lines, passing above the FR, are stretched upward, while their bottom parts are pushed against each other below the FR  
in order to fill the region where the FR was present earlier on. As a result, a current sheet is created 
behind the FR. This induces magnetic reconnection, which results in the formation 
of closed field lines at low heights (flare loops) as well as a further build-up of the erupting FR, by creating new twisted field lines wrapping around the original FR. 
This reconnection is crucial as it allows the erupting FR to be ejected toward the interplanetary space as a CME. However, even with the fastest possible reconnection 
rate, the entire incoming magnetic flux below the FR cannot be reconnected,
thus a long and thin current sheet was predicted behind an erupting FR \citep{Lin2000}.
Indeed, pieces of evidence of the current sheet formation during solar eruptions were 
reported \citep[\eg ][]{Takasao2012, Innes2015, Scott2016, Cheng2018, Lee2020}. 

Observations reveal another aspect of eruptions with contracting and expanding 
loops that were observed during eruptive solar flares 
\citep[][and references cited therein]{Veronig2006, Joshi2007, Liu2009, Liu2012, Simoes2013, 
Kushwaha2014, Petrie2016, Dudik2016, Dudik2017, Dudik2019}. 
\citet{Liu2009} and \citet{Joshi2009} found that the coronal loop contraction is associated with the converging motion of the conjugate hard X-ray (HXR) footpoints and the downward motion of the HXR loop top sources. 
The contracting and expending coronal loops are located at the 
periphery of the legs
of the erupting FR, then these loops are different than the flare loops formed by reconnection behind the erupting FR. Also, the loops which are located near the active region (AR) contract first, whereas the loops located far away contract later on \citep{Gosain2012,Simoes2013,Shen2014}.
Both \citet{Shen2014} and \citet{Kushwaha2015} observed the contraction of loops in the pre-flare phase for a duration of about half an hour.
\citet{Dudik2016} also reported the expansion and contraction of loops during an X-class flare. They interpreted this as the result of growing FR that subsequently erupts.
Next, they presented the observations
of loop expansion and contraction for two eruptive flares
(one major GOES X-class and another small GOES C-class).
In these events, the expanding and contracting loops coexist for a period of more than half an hour.
The observed speeds varies from 1.5 to 39 \kms. 
\citet{Wang2018} considered four events with loop contraction.
In the first event, the contraction occurred during the impulsive phase.
For the second event, the prominence was already erupting before the contraction (see their movie).
Finally, for the two last events, no evidence of ejection was present. 

In earlier papers, the loop contraction was associated with the conjecture proposed by \citet{Hudson2000}. According to this conjecture in low plasma $\beta$ and with negligible gravity, the magnetic energy released in a solar eruption must originate in a ``magnetic implosion". More specifically, some portion of solar corona needs to implode in order to decrease the total magnetic energy, then to a power a solar eruption. 
\citet{Shen2014} and \citet{Russell2015} instead interpreted the loop contraction
as a modification to the equilibrium of nearby loops due to the balance between the magnetic pressure and the magnetic tension.
\citet{Liu2009b} also reported the expansion and contraction of some coronal loops.  They found the expansion and contraction of the overlying coronal loops during the eruption, such that when the filament started to rise, the loops were also pushed upward, and as soon as the filament rose explosively, they began to contract. In the slow phase of the eruption, loops expanded at a slow speed ($\sim$8 \kms). 
This expansion became fast in the impulsive phase, with a speed roughly equal to the speed of the eruption ($\sim$56 \kms). When the filament erupted out, the loops began to contract inward with a speed ranging from 60 \kms\ to 140 \kms.

In contrast, \citet{Dudik2016} proposed that the apparent implosion is a result of the large-scale dynamics involving the FR eruption in three dimensions. 
Next, \citet{Dudik2017} explained their observations of two eruptive flares on the basis of the three-dimensional (3D) MHD model of erupting FR proposed by \citet{Zuccarello2017}. 
This model was derived from the analysis of 3D line-tied visco-resistive MHD simulations realized with the OHM-MPI code \citep{Zuccarello2015}. 
In this model, the eruption is triggered by the torus instability and the corona is treated as a zero $\beta$ plasma without gravity. According to these numerical simulations, the coronal loop expansion and contraction during the FR eruption are the result of hydromagnetic effects related to the generation of a vortex on each side of the erupting FR.
These vortices lead to advection of closed coronal loops.  Outward flows and returning flows are in the vortex part close and further away from the erupting FR, respectively. 
This implies the possible simultaneous presence of loops in expansion and contraction.

In this paper, we analyze the observations of the prominence eruption on 02 March 2015.  This event shows the erupting FR and the associated flare loops in various EUV and X-ray wavelengths. 
Moreover, coronal loop expansion and contraction are well observed on one side of the eruption. The paper is organized as follows:
Sect.~\ref{sect_Observations} presents the observational data, with a description of the AR magnetic field evolution on the solar disk then of the temporal and spatial evolution of the prominence eruption.  Next, the time-distance analysis of loop expansion and contraction is presented in Sect.~\ref{sect_Loop}, followed by a theoretical analysis. 
Finally, we summarize our results and conclude in Sect.~\ref{sect_Conclusion}.

\begin{figure*}     
\centering
\includegraphics[width=0.75\textwidth]{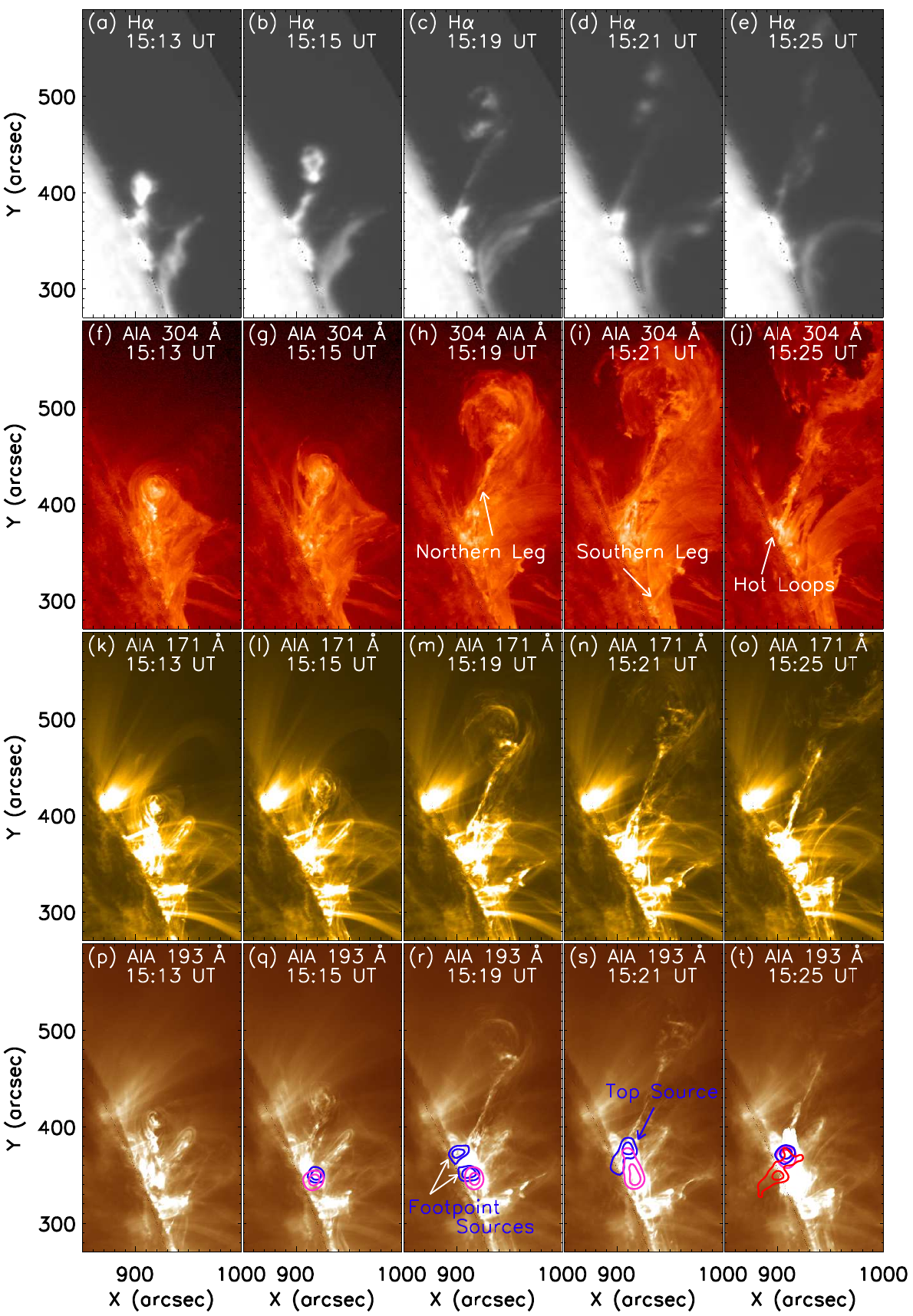}
\caption{Spatial evolution of the prominence eruption observed in GONG \ha\ (a -- e), AIA 304 (f -- j), 171 (k -- o), and 193 (p -- t) \AA\ wavelengths (see Electronic Supplementary Materials for the movies).
The onset of the eruption is $\approx$ 14:40 UT.
The northern and southern legs of the filament are labeled as the ``Northern Leg'' and 
``Southern Leg'' in panels (h) and (i), respectively.
RHESSI X-ray contours of 6 -- 12 (pink), 25 -- 50 (blue), and 50 -- 100 keV (red) energy ranges were over-plotted on AIA 193 \AA\ images in (q -- t).
The contour levels were set to 50\% and 80\% of the peak flux of the X-ray flux peak. The integration time for RHESSI images is 20 sec.}
\label{fig_eruption_evolution}
\end{figure*}

\begin{figure*}     
\centering
\includegraphics[width=0.8
\textwidth]{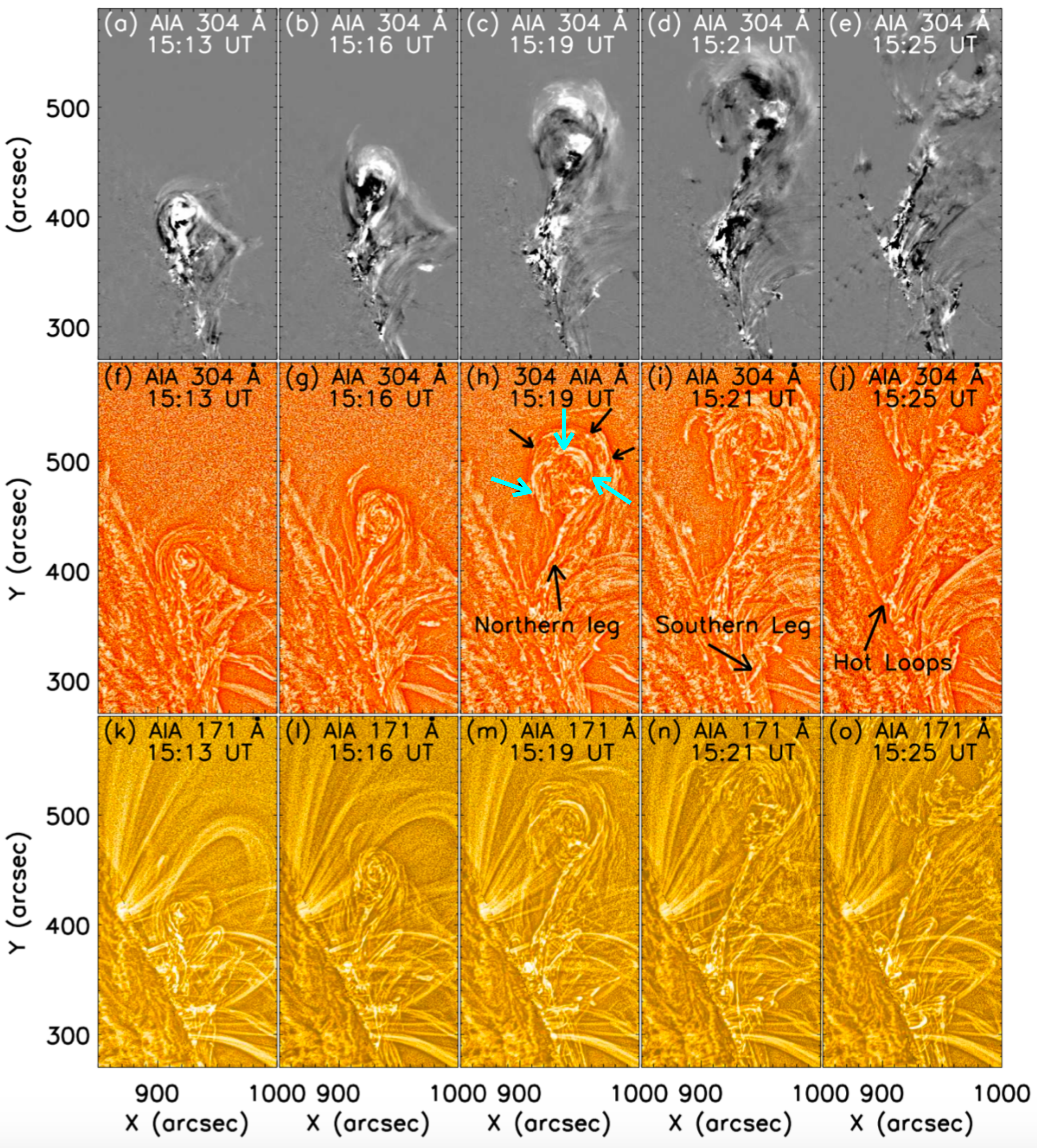}
\caption{Spatial evolution of the eruption using the running difference method (a -- e)
and with MGN processed images of the prominence eruption in AIA 304 (f -- j) and
171 (k -- o) \AA\ wavelengths (see Sect. 2.1). The northern and southern legs of the filament are indicated by arrows in panels (h) and (i), respectively.
The flare loops represented by ``hot loops'' are shown in panel (j). The black and cyan arrows in panel (h) show the outer and inner quasi circular features which is the FR configuration.}
\label{fig_eruption_evolution1}
\end{figure*}

\section{Observations}
\label{sect_Observations}

On 02 March 2015, an M3.7 class flare, associated with a prominence eruption, occurred between 15:10 UT and 15:40 UT in the AR NOAA 12290, close to the western limb of the Sun (N21W86).
For our present study, we have analyzed the data from the following instruments:

Firstly, the \emph{Atmospheric Imaging Assembly} \citep[AIA: ][]{Lemen12}
on board the \emph{Solar Dynamics Observatory\/} \citep[SDO: ][]{Pesnell12} observes the different layers of the Sun in seven EUV channels, two UV channels, and one white-light channel with a cadence of 12 seconds, 24 seconds, and 3600 seconds, respectively. 
The pixel size of AIA data is 0$''$.6. In this paper, we have analyzed AIA images in 304 \AA, 171 \AA, and 193 \AA\ wavelengths to study the prominence eruption, flare, and loop expansion and contraction during the eruption. The AR magnetic configuration was analyzed with the  
\emph{Helioseismic Magnetic Imager} \citep[HMI: ][]{Schou12}. The cadence of HMI data 
is 45 seconds and the pixel resolution is 0$''$.5.\

Secondly, the \emph{Reuven Ramaty High Energy Solar Spectroscopic Imager\/} \citep[RHESSI: ][]{Lin02} observed this flare from the rise phase at 15:15 UT until its decay phase at 15:35 UT.
RHESSI observes the full Sun with an unprecedented combination of spatial resolution
(as fine as $\approx$ 2$''$.3) and energy resolution (1 -- 5~keV) in the energy range from 3~keV to 17~MeV. \

Thirdly, the eruption was also observed by the \emph{Global Oscillation Network Group\/}
        \citep[GONG:][]{Harvey2011} in the \ha\ center.  
        GONG continuously observes the full Sun with observatories spread around Earth.
The temporal and the pixel resolution of this data are 1 minute and 1$''$, respectively.

The morphology and the temporal evolution of the prominence eruption and of the associated flare are observed in \ha, EUV, and X-rays. They are described in the following subsections.

\begin{figure*}  
\sidecaption
\centering
\includegraphics[width=0.7\textwidth]{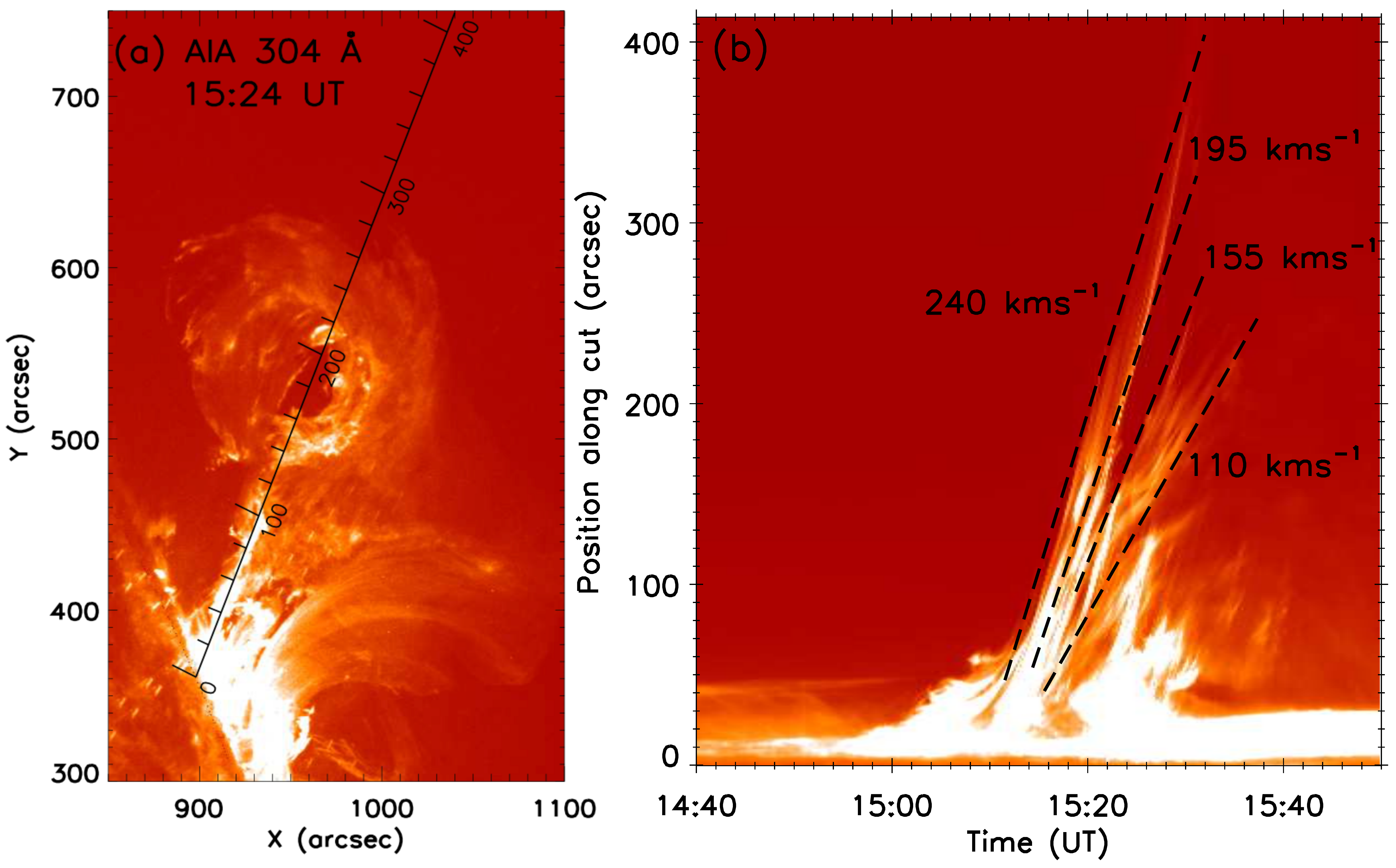}
\caption{(a) AIA 304 \AA\ image showing the ejected plasma  
 and the position of the selected slice. (b) Time-distance  
plot along the selected slice. When the emission is less saturated and the spatial extension large enough, the erupting plasma has a range of velocities as
displayed with over-plotted straight lines approximating the 
mean plasma blob trajectories (velocities are still increasing with time). This outlines the global expansion of the FR.}
\label{fig_ht}
\end{figure*}

\subsection{Morphology of the prominence eruption}
\label{sect_Morphology}

Figure \ref{fig_eruption_evolution} displays the evolution of the prominence eruption observed in \ha, AIA 304, 171, and 193 \AA. 
The onset of the eruption is $\approx$ 14:40 UT and almost simultaneous in all wavelengths (see the movie attached to Fig. \ref{fig_eruption_evolution} and the analysis in Sect. \ref{sect_Loop_Obs}). 
The major part of the prominence erupts non-radially in the northwest direction.
The extension of the erupting structure is better outlined in 304 \AA\ with a distribution of emitting plasma showing a quasi circular type of feature.
 The filament height 
grows with time and a void of emitting plasma in 304 \AA\ develops, outlining the core of the FR (Fig. \ref{fig_eruption_evolution}i,j). The quasi circular pattern and the central void are indications that the line of sight is nearly aligned with the local direction of the FR axis. This event is comparable to the one studied by \citet{Shen2014}; however, it is more clearly structured than the other event, likely because it was observed more along the FR axis direction.

During the eruption, some part of the prominence is diverted toward the west side 
(Fig. \ref{fig_eruption_evolution}g -- j and the attached movie).  We interpret this as prominence plasma which is initially located southward and within a part of the FR more inclined on the line of sight. Then, the FR axis changes along the configuration which is a 3D writhed configuration.  The presence of a twisted field is much less obvious in this southern part as the FR is seen partly from the side.
 
During the uplifting of the filament, the coronal plasma shows a twisted structure in AIA 171 and 193 \AA\ (Fig. \ref{fig_eruption_evolution} bottom rows).
However, we cannot define a sign for the twist because we cannot determine which structures are in the foreground and background with the optically thin coronal emission.
The presence of a prominence before the eruption is compatible with the FR since an upward curvature of magnetic field lines is needed to support the dense prominence plasma against gravity. However, a sheared magnetic arcade can also possess magnetic dips and hence be capable of supporting a filament against gravity \citep[see the review of ][]{Mackay2010}.

In \ha\ observations, the upper part of the prominence is first seen to get detached  from a remaining part staying above the chromosphere (Fig. \ref{fig_eruption_evolution}a,b).
Later on, the erupting prominence splits into two parts (Fig. \ref{fig_eruption_evolution}c,d).  
The superposition with the co-aligned 304 \AA\ data shows that one part is located below the FR center, which is as expected in the gravitationally stable configuration with upward curvature of the field lines, while the other part is located well above the FR axis which is not expected.  
This implies that enough kinetic energy should be injected into this plasma during the earlier phase of eruption in order to bring it up in the FR.  
This observation could be related to the results of \citet{Lepri2010} at 1 AU, which were obtained using the Solar Wind Ion Composition Spectrometer (SWICS) measurements.  They found that cold 
plasma could be present in a fraction of the extension of some magnetic clouds, with no 
specific  location, in particular not necessarily in the rear part.

As the prominence erupted, a straight and emitting structure is formed behind the erupting FR (Fig. \ref{fig_eruption_evolution}h).
This is most likely the northern filament leg that is stretched thin and long by the erupting magnetic configuration. Next, this structure gets broader and less coherent with time (Fig. \ref{fig_eruption_evolution}i,j).
The southern filament leg is also visible in Fig. \ref{fig_eruption_evolution}i.
In the later phase of the eruption, flare loops formed which are seen in projection below the northern leg of the filament (see Fig. \ref{fig_eruption_evolution}j), although we cannot say it is actually below the leg of the filament or not because of the line-of-sight confusion on the limb.

For a better understanding of the eruption, we created images using the running difference method (Fig. \ref{fig_eruption_evolution1}, top row). These images were obtained by the subtraction of the image obtained one minute before.  
These images outline the development of the filament eruption in AIA 304 \AA\ with a contrast better than the original images as the method emphasizes the changes.  The FR is better seen in the difference images at 15:19 and 15:21 UT with dark gray and quasi circular-like structures (Fig. \ref{fig_eruption_evolution1}c,d). 

A limitation of the running difference method is that we do not see the loops well.
Then, we also applied the Multi-Gaussian Normalization (MGN) method developed by \citet{Morgan14}.
The evolution of the MGN images of AIA 304 and 171 \AA\ are presented in the middle and bottom rows of Fig. \ref{fig_eruption_evolution1}, respectively.
The MGN method defines a normalized image by using the local mean and standard  deviation computed with a local Gaussian function (called a kernel).
The observed image was first convolved with the kernel to compute the local mean and standard deviation.  
By subtracting the local mean and dividing by the local standard deviation, the MGN method defines the normalized image. Then, this normalized image is transformed by the arctan function. 
This process was repeated with different kernel widths, and the final image is a weighted  combination of the normalized components. 

\begin{figure*}     
\sidecaption\centering
\includegraphics[width= 0.7\textwidth]{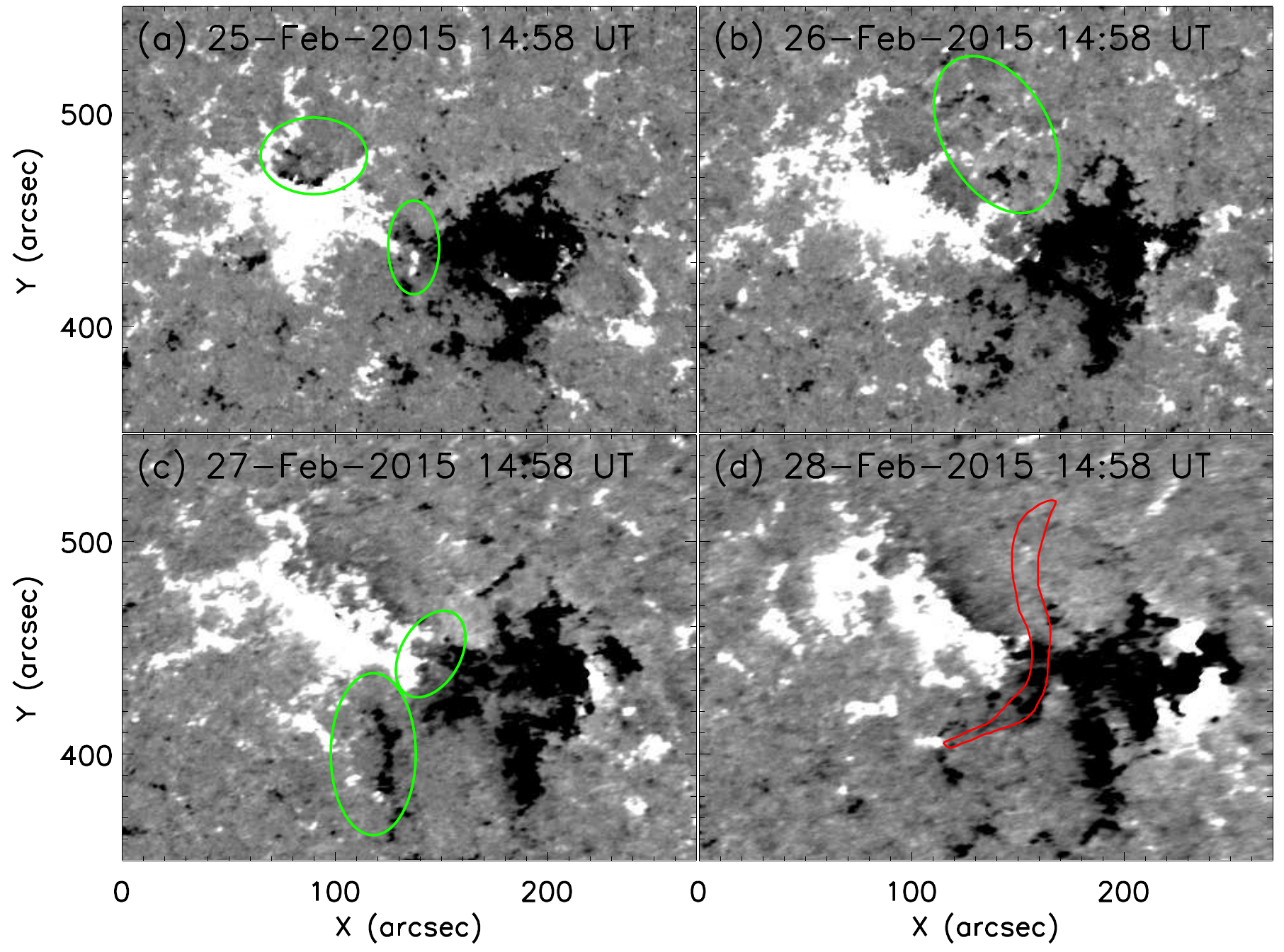}
\caption{Evolution of the AR magnetic field from 25 to 28 February 2015.
The locations of the flux cancellation are shown by green ellipses.  The location of the \ha\ filament on 28 February 2015 is displayed in panel (d) with a red contour.
A movie of magnetic field during 25 -- 28 is available in the Electronic Supplementary Materials.}
\label{fig_HMI}
\end{figure*}

Compared to Fig. \ref{fig_eruption_evolution}, the MGN method emphasizes the northern and southern legs of the filament as pointed in Fig. \ref{fig_eruption_evolution1}h,i. The internal structure of the erupting FR is also better seen than in the original images.  When well identified with the MGN method, most of the structures can also be identified in the original images, albeit less precisely. In particular, parts of circular structures are also present inside the quasi circular structure identified in Fig. \ref{fig_eruption_evolution} and marked with arrows in Fig. \ref{fig_eruption_evolution1}h.   With a likely 3D writhed configuration outlined by only a few loops that are dense enough, it is indeed remarkable to be able to see the traces of an FR in part of the erupting configuration.
Next, similar erupting structures are present in the different AIA filters.
This indicates an important plasma component within the range of transition region temperatures
because this is the temperature range shared by the temperature 
response functions of the three AIA EUV filters.

The MGN method also allows one to see the loops surrounding the erupting FR, since they are barely seen in the original data.  These loops are not necessarily near the eruption as the integration length along the line of sight is large at the solar limb. 
For example, the eruption is seen to progress on a foreground or background of undisturbed 
large-scale loops in AIA 171 \AA , while other loops, of a similar height as the eruption in Fig. \ref{fig_eruption_evolution1}l,m, are evolving; the optically thin emission did not allow us to determine which structure is in front. The full dynamic of the coronal loops and of the eruption is shown in the associated movies.   

The prominence plasma is nearly at a constant location before 14:20 UT and later on it progressively accelerates outward with a significant inclination on the local vertical (Fig. \ref{fig_ht}).  The early part of the eruption, before 15:15 UT, is too saturated in 304 \AA\ to see more than a global upward motion.  Later on, an increasing spreading of the erupting plasma 
started to develop
with different plasma blobs moving at different speeds.
The average speeds range from 110 to 240 \kms\ within the FR. 
These speeds were calculated using a time distance analysis of AIA 304 \AA\ data sets. In Fig. \ref{fig_ht}b, we outline the main structures with straight dashed lines.
Plasma in different parts of the FR move at a different speed. We interpret the range of speed to be due to the expansion of the FR.   The cone shape defined by these lines indicates that the FR expands nearly self-similarly with a linear rescaling with time of its spatial configuration (\ie\ the FR is rescaled globally with a scalar factor depending linearly on time).  
The expansion speed of the FR border, relative to the FR center, is $\approx$ 65 \kms, which is
slightly more than one-third of the speed of the FR center ($\approx$ 175 \kms), then we conclude that the expansion is significant.

Later on, this prominence eruption was accompanied by a CME observed by the {\it Large Angle and Spectrometric Coronagraph} (LASCO, \citet{Brueckner95})
onboard the SOHO satellite and it is given in the SOHO/LASCO CME catalog at  
https://cdaw.gsfc.nasa.gov/CME$\_$list/ \citep{Yashiro2004}. The CME first appears in the LASCO C2 field-of-view at $\approx$ 15:45 UT. The speed of the CME leading edge is 452 \kms\ so it is less than twice its velocity in the low corona (Fig. \ref{fig_ht}b). At the distances observed by C2, the CME was already decelerating with a mean value of deceleration of 12 m~s$^{-2}$.  

\begin{figure*}     
\centering
\includegraphics[width=\textwidth]{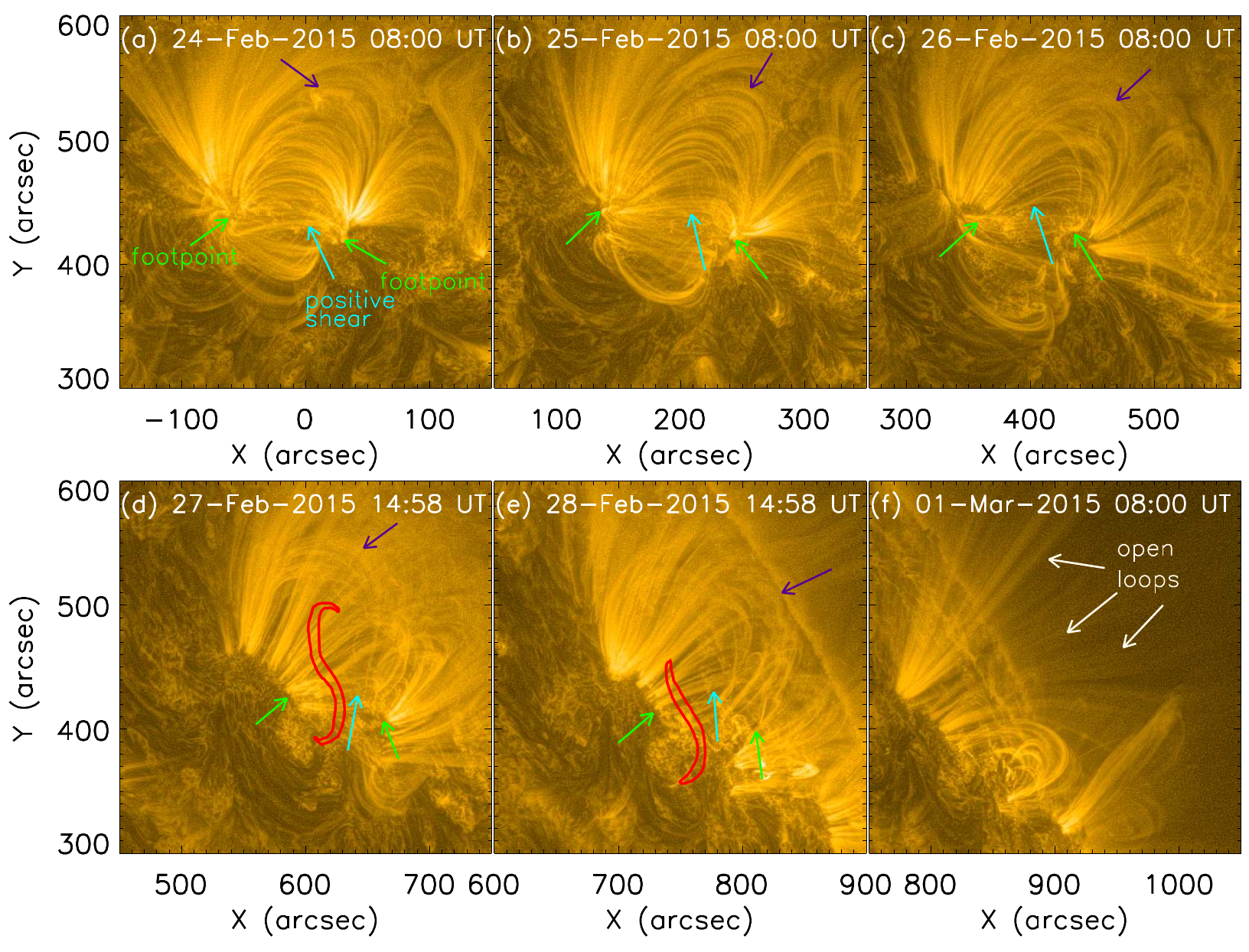}
\caption{Evolution of the corona in AIA 171 \AA\ from 6 to 1 day before the studied eruption. A sheared loop built up in the AR core from 24 to 28 February (green and blue arrows in panels (a -- e). The purple arrows point to large scale loops of the AR (a -- e) which become open, or at least large scale in panel (f). The location of the filament is displayed in panel (d, e) with a red contour. 
}
\label{fig_evolution_171}
\end{figure*}

\subsection{Photospheric magnetic field}
\label{sect_Photospheric_B}

We analyze the AR magnetic configuration with the HMI magnetic field data.  Since the studied eruption was at the limb on 02 March 2015, the photospheric magnetic field is not available. 
Therefore, we follow its associated AR evolution when it was on the solar disk.  
AR NOAA 12290 was on the eastern limb on 18 February 2015.  One day later, it started to show on the solar disk as a small bipolar AR with a negative leading and a positive following polarity.  During its disk crossing, it progressively grew in size as its magnetic field got dispersed by convective motions. Then, AR 12290 was well in its decay phase when it reached the western limb.

Figure \ref{fig_HMI} shows four magnetograms of the longitudinal field component with a time cadence of one day.  
The AR was located at $\approx$ 15$\degree$ of longitude after the central meridian on 25 February 2015 in panel (a), and at $\approx$ 55$\degree$ of longitude on 28 February 2015 in panel (d). The magnetograms are spatially de-rotated to the central meridian so that the spatial extension of the polarities could be compared between the different panels.   However, we kept the longitudinal component so projection effects appeared as the AR 
approached the limb (mostly on the limb side of the leading polarity where fake positive polarities appear).
The magnetic field mostly disperses with time in each polarity so that 
the polarities become more extended. Then, the global bipolar magnetic configuration is progressively larger in size, which is expected to induce a global expansion of the coronal loops. This dispersion also implies that the polarities of opposite signs become into contact at the PIL.
This induces cancellations of the magnetic flux, for example,\ at the locations surrounded by a green ellipse in Fig. \ref{fig_HMI}a,c. 
These cancellations of flux are best seen in the movie associated with Fig. \ref{fig_HMI}. 
Such cancellations imply a progressive buildup of an FR above the 
PIL \citep{Ballegooijen1989,Amari2003,Green2018}. 
Finally, the FR becomes unstable leading to an eruption (Sect. \ref{sect_Morphology}).  

The above analysis shows that AR 12290 is mainly a decaying bipolar AR with cancellations occurring at the PIL.  This evolution is expected to continue slowly in the following days as typically observed in decaying ARs \citep[see the review of][]{vanDriel-Gesztelyi2015}.  
Therefore, it is justified to take the closest possible on-disk magnetic field of the western limb to approximate the photospheric field distribution during the eruption.  
The last magnetogram shown (Fig. \ref{fig_HMI}d) is about 2 days before the studied eruption at the limb.  The location of the filament on 28 February 2015 is over-plotted with a red contour. 

Next, AR 12290 has the typical configuration of ARs with a more dispersed
following polarity than the preceding one \citep{vanDriel-Gesztelyi2015}.
This may lead to the non-radial eruption as present in numerical simulations \citep{Aulanier2010}.
However, the asymmetry seen in Fig. \ref{fig_HMI} is moderate and the tilt of the AR bipole on the solar equator is small while the observed eruption is tilted well away from the radial direction (Fig. \ref{fig_eruption_evolution}).

\begin{figure}    
\centering
\includegraphics[width=0.5\textwidth]{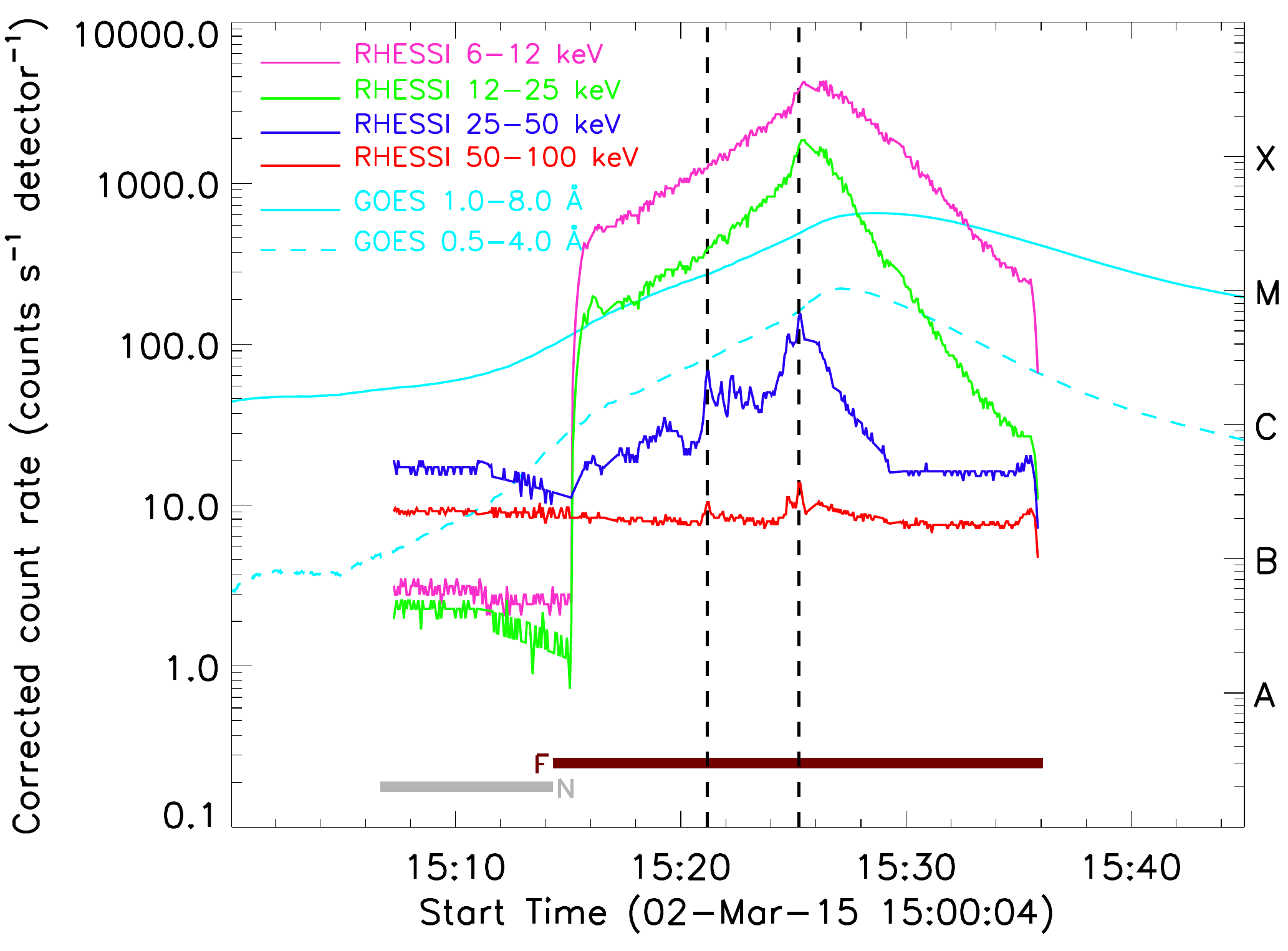}
\caption{GOES and RHESSI X-ray time profiles of the M3.7 flare. For RHESSI, profiles were
plotted in the energy bands of 6 -- 12 keV (pink), 12 -- 25 keV (green), 25 -- 50 keV (blue), and 50 -- 100 keV (red). For clarity of presentation, we scaled RHESSI count rates by factors of 1, 1/2, 1, and 1/5 for 6 -- 12 keV, 12 -- 25 keV, 25 -- 50 keV, and 50 -- 100 keV energy bands, respectively. RHESSI light curves were corrected for change in attenuator states during flare observations.
Horizontal bars at the bottom represent the RHESSI observing state (N: night; F: flare).
}
\label{fig_rhessi_lc}
\end{figure}

Fig. \ref{fig_HMI}, and more precisely the associated movie, shows the convergence of polarities of both signs at the PIL. This induces a magnetic shear and indeed positively sheared loops are observed to develop from 24 to 28 February with AIA 171 \AA\ observations  (Fig. \ref{fig_evolution_171}a -- e).
The polarity cancellation is expected to transform these sheared loops in an FR. This cancellation happens, in particular, at the northern part of the PIL (Fig. \ref{fig_HMI}a -- c). A C3.7 flare starts after 05:00 UT on 01 March 2015.  This is the largest event observed in AR 12290 since the beginning of its disk transit. This event is eruptive and associated with a slow CME with a speed $\approx$ 191 \kms.
This event changes the coronal configuration drastically by opening the magnetic field on the northern part of the AR.  
This is seen after the flare in AIA 171 \AA\ observations by the disappearance of the coronal loops, and the creation of a region of low emission embedded in open, or at least large-scale, loops (Fig. \ref{fig_evolution_171}f). 
Then, we interpret the non-radial motion of the filament as a consequence of the transformation generated by the previous event, which occurred around 05:00 UT on 01 March 2015.  This event opens, or at least creates larger scales, within the northern coronal field of the AR. This new configuration is expected to channel the later filament eruption, which is then diverted from the central AR part toward its northern side (Fig. \ref{fig_eruption_evolution}).

\subsection{Temporal and spatial evolution of X-ray sources} 
\label{sect_Flare}
The physical processes during a flare associated with the prominence eruption, in particular the transformation of magnetic to kinetic energy (acceleration of particles), are best characterized by X-ray emissions.
The temporal evolution of the flare in HXR and associated sources yield insights about the eruption phenomena (heating and nonthermal emissions) in the source region.
For this purpose, we compare GOES and RHESSI light curves for the M3.7 flare associated with the prominence eruption in Fig. \ref{fig_rhessi_lc}.
The GOES 1 -- 8 \AA\ time profile shows a typical temporal behavior of a long duration event during $\approx$ 15:00 -- 15:40 UT with a distinct peak at $\approx$ 15:28 UT. 
The peak in a higher energy light curve of GOES ({\it i.e.}, 0.5 -- 4.0~\AA) occurs slightly earlier at $\approx$ 15:27 UT, which is expected.
RHESSI profiles at energies $\leq 25$~keV indicate a continuous rise in the X-ray count rate until $\approx$ 15:26 UT  and they decline thereafter. 
From these RHESSI light curves, we note the rise phase to be more gradual than the decline phase which is rather unusual.

RHESSI 25 -- 50 and 50 -- 100 keV light curves  
have a maximum at $\approx$ 15:25 UT.
Notably, in the 25 -- 50 keV profile, we observe significant fluctuations with a few pronounced sub-peaks during the rise phase.
Furthermore, a prominent sub-peak around 15:21 UT can also be seen at the high energy channel of 50 -- 100 keV (indicated by left vertical line).
These sub-peaks at $>$25 keV represent distinct episodes of particle acceleration as the erupting FR forces magnetic reconnection events underneath it.
Apart from two brief intervals (around the two vertical lines), the flare does not show flux enhancement in the 50 -- 100 keV energy band observations.

\begin{figure*}[t!]     
\sidecaption
\centering
\includegraphics[width=0.7\textwidth]{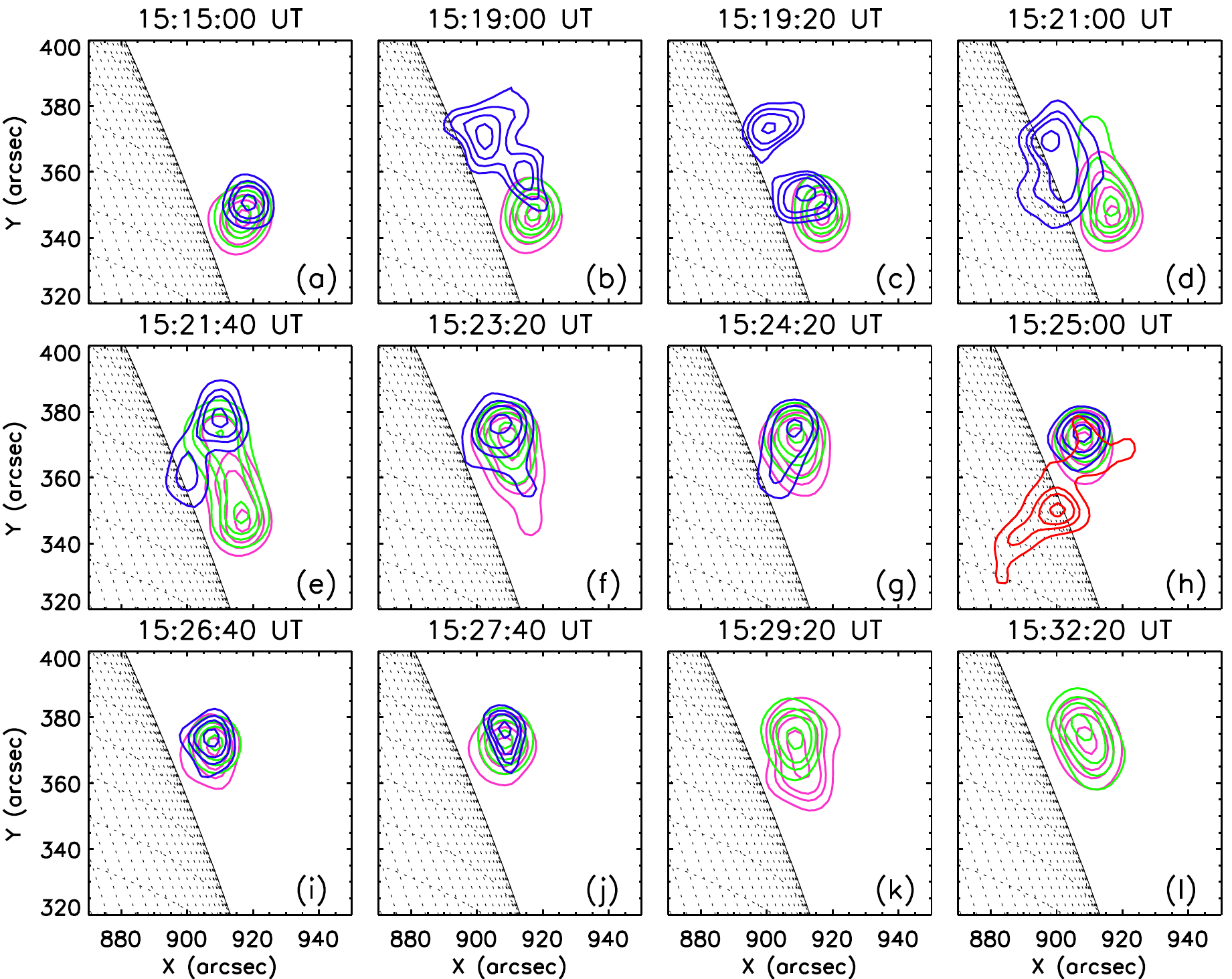}
\caption{A few representative RHESSI intensity isocontours showing relative positions and spatial
evolution of X-ray sources in 6 -- 12 keV (pink), 12 -- 25 keV (green), 25 -- 50 keV (blue), and
50 -- 100 keV (red) during the M3.7 flare.  The contour levels were set at 55\%, 70\%, 80\%, and 95\% of the peak flux of each image.
The integration time for each image is 20 sec.}

\label{fig_rhessi_sources}
\end{figure*}

Limb events provide us with a unique opportunity to distinguish the HXR emission
that originated at the coronal loop tops from that of their foot-points.
To discuss this in detail and to study the morphological evolution of HXR sources,
we utilized the imaging capabilities of RHESSI. We reconstructed RHESSI images by the CLEAN algorithm with the natural weighing scheme \citep{Hurford2002} using front detector segments 3 -- 8 (excluding 7).
The images are produced at 6 -- 12, 12 -- 25, 25 -- 50, and 50 -- 100 keV energy bands.
In Fig. \ref{fig_rhessi_sources}, we present a series of co-temporal RHESSI images in different energy channels.
The counts statistics during the selected times, as seen from the time profiles, together with the integration time of 20 sec ensure the reliability of the HXR source structures.
To examine the spatial location of HXR sources with respect to erupting prominence, we plotted HXR contours over AIA 193 \AA\ images in Fig. \ref{fig_eruption_evolution}q -- t.

Comparisons of RHESSI images at different channels show a complex structure and evolution of the X-ray sources. At the beginning, the X-ray sources, which were observed up to 50 keV energies, are co-spatial (Fig. \ref{fig_rhessi_sources}a). This X-ray emitting region lies in the lower corona where the initiation of the plasma eruption occurred. Next, we note the onset of the X-ray emission from a new location which lies northeast of the initial X-ray emitting region (Figs. \ref{fig_eruption_evolution}r and \ref{fig_rhessi_sources}b,c).
Around the first HXR peak ($\approx$ 15:21~UT), the X-ray sources at 25 -- 50~keV exhibit an extended structure with a distinct appearance of a coronal source beside another one at lower heights (Fig. \ref{fig_rhessi_sources}e). In the subsequent phases, the X-ray emission is only observed  from the latter, which developed in the northeastern X-ray emitting site (Fig. \ref{fig_rhessi_sources}f -- l). 
It is noteworthy that, although the 50 -- 100~keV sources appear very briefly during the flare maximum at 15:25 UT, they exhibit an extended structure from lower to higher coronal heights with the strongest source extending onto the solar disk (Fig. \ref{fig_rhessi_sources}h).
These sources probably are a composite emission from the coronal and foot-point regions (see also Fig. \ref{fig_eruption_evolution}t). 
Finally, the coronal emission continued during the decline phase at energies up to 25 keV (Fig. \ref{fig_rhessi_sources}i -- l).   

The temporal and imaging analyses of HXR emission suggest a large variability in the flux for all sources is present, especially in the hardest channels. 
These observations are in agreement with the standard model of the formation of a coronal source above the flare loops observed in EUV and two foot-point sources in the magnetically connected foot-points.
These X-ray sources are expected to be formed by the energetic particles that accelerate at the base of the reconnecting region or in the newly formed flare loops, which are typically in contraction, and thus a favorable configuration for particle acceleration \citep[\eg\ see the review by][]{Benz2017}.

\begin{figure*}[t!]     
\centering
\includegraphics[width=0.93\textwidth]{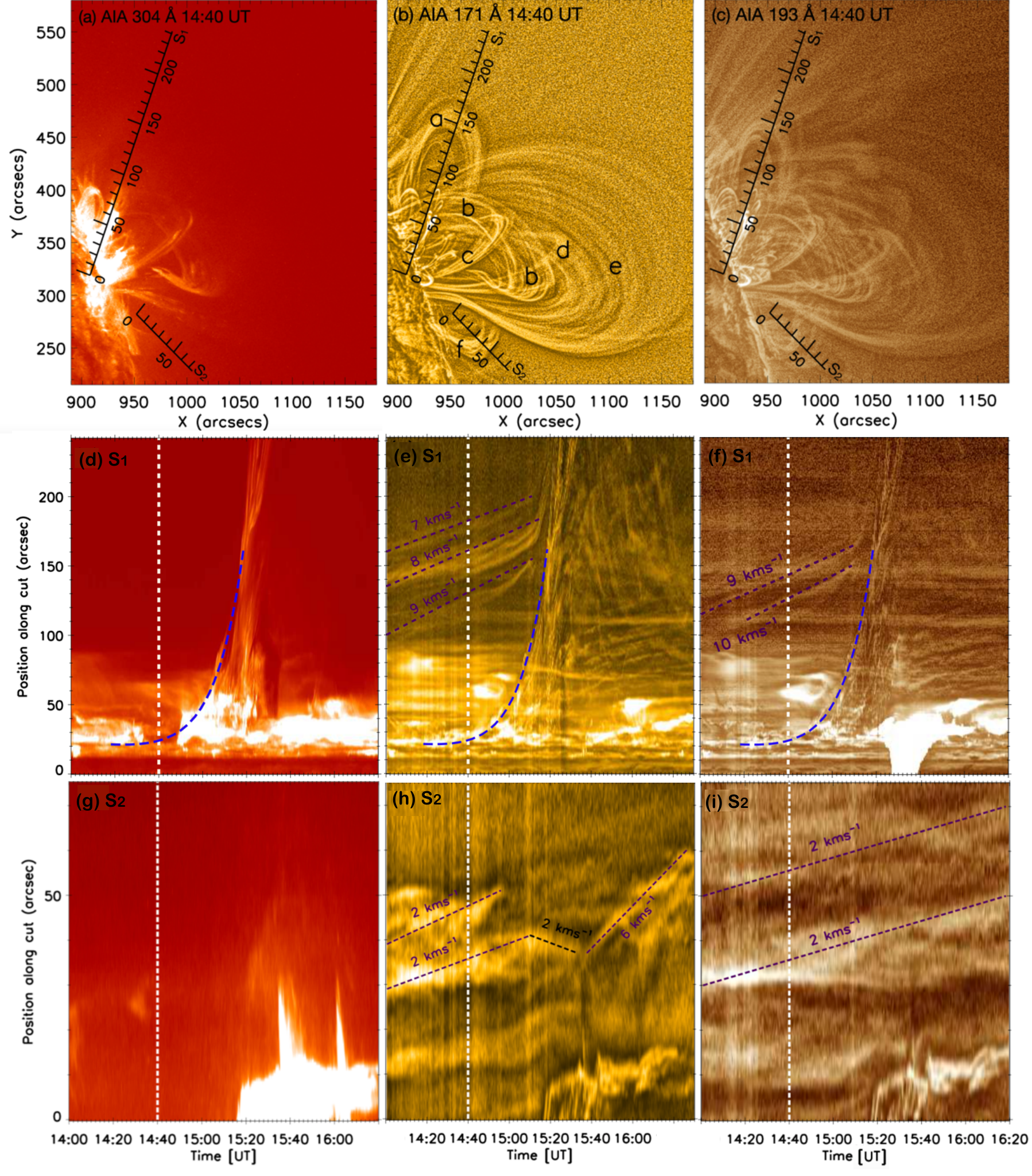}
\caption{
Location of slices S$_1$ and S$_2$ (top row) chosen for the time-distance analysis of AIA 304 (a), 171 (b), and 193 (c) \AA. The time-distance plots for slice S$_1$ and S$_2$ are presented in the middle and the bottom rows, respectively, in AIA 304, 171, and 193 \AA. The blue dashed line in panels (d-f) is the fit of a combination of a linear and exponential function to the trajectory shown in the time slice data.
The vertical white dashed line in the time-distance plots indicates the onset time of the prominence eruption and also the time of the images shown in the top row.
Slice S$_1$ was selected in the direction of the prominence eruption so that the erupting plasma would be clearly visible in the time-distance image (middle row). 
Slice S$_2$ was selected on the opposite side of the AR from the eruption region to explore the extension of the eruption effects. 
A movie of AIA 304, 171, and 193 \AA\ is available in the Electronic Supplementary Materials.}  

\label{fig_timeslice_s1_s2}
\end{figure*}

\section{Loop expansion and contraction} 
\label{sect_Loop}

\subsection{Observational evidences}
\label{sect_Loop_Obs}

In the present study, we found observations of coronal loop expansion and contraction at various EUV wavelengths of the AIA telescope during the prominence eruption on 02 March 2015. 
To understand this phenomena, we created time-distance plots in several directions and finalized four slices. 
We name these slices S$_1$, S$_2$, S$_3$, and S$_4$.  
The measured speeds for all the selected loop systems and the time ranges for the expansion and contraction are tabulated in Appendix \ref{sect_Measured_Velocities} with Table \ref{tab_velocity}.

\smallskip   
\noindent {\bf Slice S$_1$:} The top panels of Fig. \ref{fig_timeslice_s1_s2} depict the location of slice S$_1$ and S$_2$. Abscissa $s_1$ and $s_2$ are defined along the cuts, and the associated time-distance plots are shown in the middle and the bottom panels, respectively. Slice S$_1$ was set in the direction of prominence eruption. From panels (e) and (f) in Fig. \ref{fig_timeslice_s1_s2}, we inferred that the eruption is initiated around 14:40 UT when a significant upward displacement of the emitting plasma could be detected after a phase with a nearly constant abscissa $s_1$. Later on, the prominence plasma progressively accelerated with an exponential behavior. Then, we used a combination of an exponential and linear increase of height as a function of time to perform a least square fit of the data \citep{Cheng2020}. The fitted function is in the form of
$f(t) = a \,e^{t/b} + c\,t +d $, where a, b, c, and d are the coefficients determined by the fit to the data. This fitted function nicely describes the prominence plasma evolution observed in 171 and 193 \AA\ with no significant difference between the two filters (panels (e) and (f) of Fig. \ref{fig_timeslice_s1_s2}).

Slice S$_1$ of AIA 304 \AA\  mostly images the plasma located at a low height around $s_1 \approx$25$''$ at 14:00 UT (Fig. \ref{fig_timeslice_s1_s2}d). This plasma becomes dark around 14:40 UT, which is the starting time of the eruption, then it becomes very bright, saturating the detector. Later on, the emitting plasma accelerated upward, up to a velocity around 200 \kms~(Fig. \ref{fig_ht}b).   
After the prominence eruption, the falling back of erupting prominence plasma was observed (see attached movie). The function fitted to 171 and 193 \AA\ data is also compatible with the 304 \AA\ observations which are less constraining at lower $s_1$ values (Fig. \ref{fig_timeslice_s1_s2}d).

We label "a" the loop system located well above the erupting prominence in the 
range of $s_1 \approx$120 -- 180$''$ at 14:40 UT (Fig. \ref{fig_timeslice_s1_s2}b).   
In both 171 and 193 \AA , the loop system "a" is in global expansion before the prominence eruption.   The expansion velocity of this loop system in AIA 171 \AA\ varies from 7 to 9 \kms , and in AIA 193 \AA\ from 9 to 10 \kms\ with a tendency of a slightly larger velocity at lower height. This expansion becomes much faster when the prominence eruption approaches $s_1 \approx$100$\arcsec$, around 15:10 UT; the prominence plasma is clearly seen in 304 \AA\ in panel (d), and at the same locations in 171 and 193 \AA\ in panels (e) and (f). 

We also find some loops in both 171 and 193 \AA\ which are stationary before and during  the prominence eruption process (\eg\ $s_1 \approx 15,\, 115,$ and $145 \arcsec$ at 14:00 UT in panels (e, f)).
These stationary structures are almost vertical (radial), and more precisely they trace the legs of large-scale loops in panels (b, c). With the projection effect at the solar limb, we observe them closer to the eruption than they really are.  These stationary loops are not likely rooted in the vicinity of the eruption site, thus they are not disturbed during the eruption process. 
 
\begin{figure*}    
\centering
\includegraphics[width=0.93\textwidth]{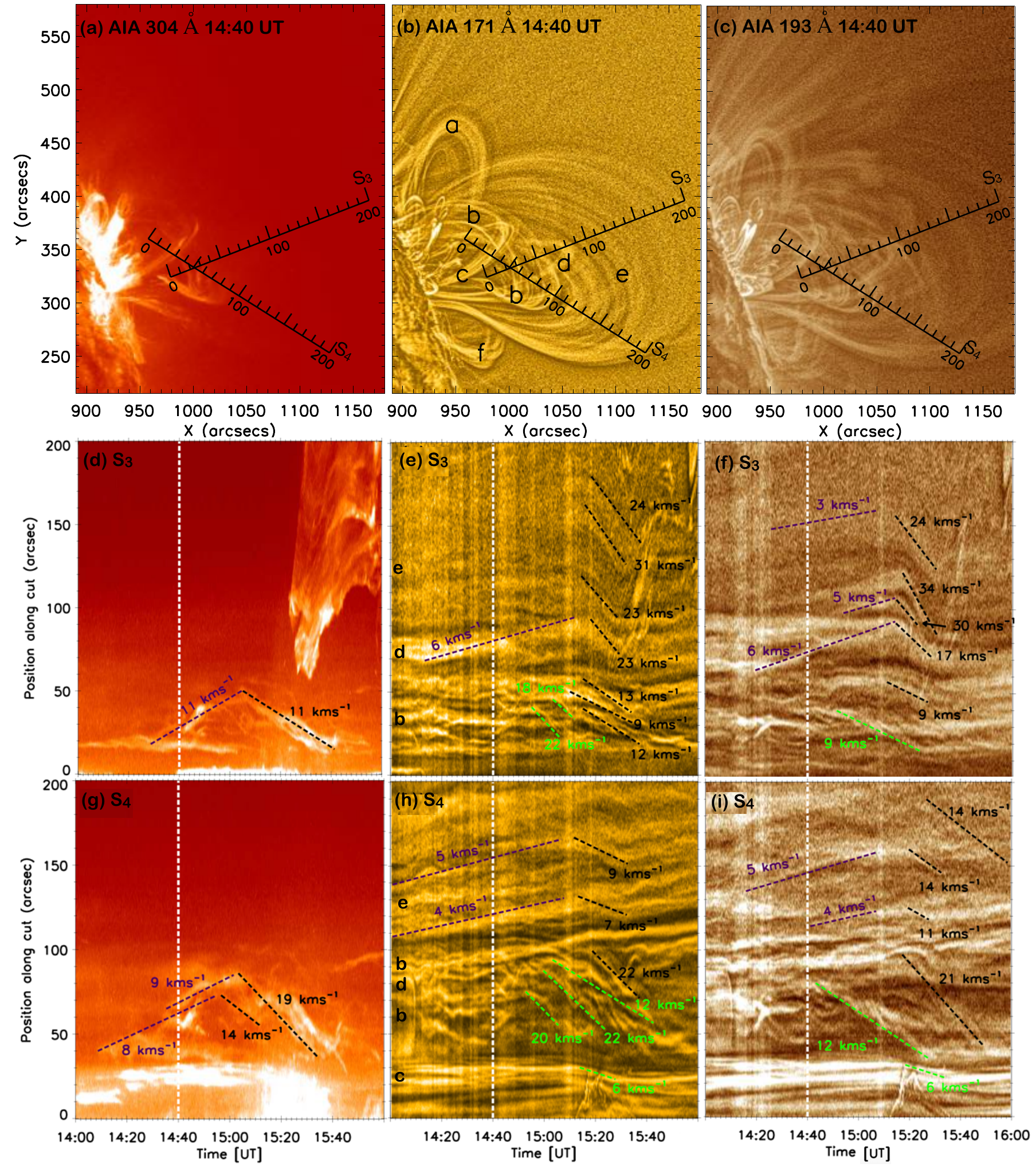}
\caption{ Location of slices S$_3$ and S$_4$ (top row) chosen for the time-distance analysis of AIA 304, 171, and 193 \AA\ presented in the panels (a), (b), and (c), respectively.
These slices were selected for the analysis of loop expansion and contraction. 
Different expanding and contracting loops are shown by dashed lines in the middle and bottom rows.
The white vertical dashed line indicates the onset time of the eruption and the time of the top row images.
The loop systems in the direction of slices S$_3$ and S$_4$ are labeled in panels (e) and (h), respectively. A movie of AIA 304 \AA\ is available in the Electronic Supplementary Materials.}

\label{fig_timeslice_s3_s4}
\end{figure*}

\smallskip  
\noindent {\bf Slice S$_2$:}
The bottom panels of Fig. \ref{fig_timeslice_s1_s2} display the time-distance plots related to slice S$_2$ in AIA 304, 171, and 193 \AA.  We label "f" the main loops in this direction (Fig. \ref{fig_timeslice_s1_s2}b). This slit tests if the eruption affects the side of the AR opposite to the eruption location. In both 171 \AA\ and 193 \AA\ wavelengths, the expansion speed before eruption is about 2 \kms , which is slow (Fig. \ref{fig_timeslice_s1_s2}h,i).
This expansion mostly continues for some loops during and after the eruption in 
193 \AA , while there is some evidences of contraction for others. The observations are clearer in 171 \AA\ with a contraction starting at $\approx$ 15:10 UT, so during the eruption, for example,\ at $s_2 \approx$ 40$\arcsec$. This is followed by a relatively fast expansion $\approx$ 6 \kms\ starting at $\approx$ 15:40 UT.

\smallskip 
\noindent {\bf Slice S$_3$:} The top panels of Fig. \ref{fig_timeslice_s3_s4} show the location of slice S$_3$ and S$_4$. The associated time-distance plots are shown in the middle and bottom panels, respectively. Before the eruption, AIA 304 \AA\ only detected plasma at $s_3 \approx$18$''$ (Fig. \ref{fig_timeslice_s3_s4}d). After $\approx$ 14:30 UT, this plasma moved up with a global expansion speed of about 11 \kms. Then, at $\approx$ 15:02 UT, the motion changed to a downward motion at a similar speed. 
At larger heights, $s_3 >50 \arcsec$, part of the ejected plasma crossed S$_3$.  

   The observations in hotter temperature ranges have different and complementary information on three main sets of coronal loops. The loops in AIA 171 and 193 \AA\ 
   crossing S$_3$ are labeled "b" ($s_3$ in 25 -- 60$\arcsec$), "d" ($s_3$ in 70 -- 85$\arcsec$), and "e" ($s_3>$ in 90$\arcsec$) at 14:40 UT (Fig. \ref{fig_timeslice_s3_s4}b). 
These loops are in global expansion with a speed in the range of 3 to 6 \kms , with the speed decreasing with height.   After this global expansion, these loops suddenly contracted at $\approx$ 15:18 UT in phase at different heights.  Later on, around 15:40 UT traces of ejected plasma were present in both 171 and 193 \AA\ while much less visible than in AIA 304 \AA\ (Fig. \ref{fig_timeslice_s3_s4}d).

A closer analysis shows that the loop systems have some differences in their evolution.
The loop system labeled "b," which is between $s_3= 25$ and $60 \arcsec$ at 14:40 UT, contracts earlier, as it starts at 14:59 UT (Figs. \ref{fig_timeslice_s3_s4}e,f).  
The speed of this earlier contraction is in the range of 9 -- 22 \kms. 
Next, the loop, labeled "d" (Fig. \ref{fig_timeslice_s3_s4}b), is likely a set of unresolved loops. 
It has a different shape and likely also different photospheric connections than loop systems "b" and "e."  Loop "d" started to contract later than "b," around 15:16 UT, in phase with all the loops located at larger heights (system "e"). 
The speed of this contraction varies from 9 to 34 \kms\ with a tendency of a larger speed at a larger height.  The magnitude of these speeds is comparable to the speed of the earlier contraction of "b," while higher than the speed of the earlier expansion.
The above speed values are coherent with the expansion and contraction speed found previously, from 
a few (1 -- 2) \kms\ to 39 \kms, for the eruption of 05 March 2012 and 19 June, 2013  \citep{Dudik2017}.

\smallskip 
\noindent {\bf Slice S$_4$:} The time-distance plots for slice S$_4$ in AIA 304, 171, and 193 \AA\ are presented in panels (g), (h), and (i), respectively, in Fig. \ref{fig_timeslice_s3_s4}. In 304 \AA,\ an upward motion, followed by a downward one, of loop system "b" is observed
as was previously observed for slice S$_3$. 
Indeed, the whole loop system "b" moves in phase (see the associate movie in 304 \AA ). 
In 171 and 193 \AA,\ the loops encountered with growing $s_4$ abscissa are labeled as "c," "b," "d," "b," and "e."
We observed a similar expansion before the eruption, followed by a contraction during the eruption as in the case of S$_3$ for the loops "b," "d," and "e." 
Again, these loops move in phase all along their length.
An earlier contraction of loops "b" and "d" is also present for $s_4<90 \arcsec$, starting at 14:59 UT just as for S$_3$ . The speed of this earlier contraction varies from 6 to 22 \kms .

\subsection{Theoretical interpretation}
\label{sect_Loop_Theoretical}
As the analyzed eruption is at the limb, it has less background and foreground than when an eruption is observed on the solar disk. Then, the early coronal evolution of this event can be observed better.  There is a weak upward motion of the prominence before the event in the time range of 14:00 -- 14:30 UT in
Fig. \ref{fig_timeslice_s1_s2}e,f. This evolution is likely driven by the photospheric cancellation of the magnetic field at the PIL,  which could only be observed days before an eruption. (Sect. \ref{sect_Photospheric_B}). Then, the prominence plasma seen in \ha\ and in AIA filters starts to progressively move up faster with an exponential growth, 
which rapidly dominates in magnitude the earlier linear evolution of its position.
The nonlinear increase in s$_{1}$ abscissa starts to be significant around 14:40 UT.  This prominence plasma traces a part of the magnetic configuration and the start of its exponential evolution indicates that the magnetic field becomes unstable. This defines the physical starting time of the eruption. 

Around the prominence eruption starting time, 14:40 UT, GOES fluxes weakly evolve with time.
This small evolution continues even much later on as both GOES channels start
to increase in flux only at $\approx$ 15:05 UT, so 25 minutes later (Fig. \ref{fig_rhessi_lc}).
The increase in EUV flux is also weak during that time period and even in localized regions of the AR. We interpret these data with the model of \citet{Lin2000}. The time interval between $\approx$ 14:40 and 15:05 UT corresponds to an almost ideal-MHD evolution of the FR which is unstable. In the model, a current sheet forms behind the FR with a reconnection rate that is too slow to process a significant amount of the magnetic field. The small amount of magnetic energy that is liberated is mostly transformed to macroscopic kinetic energy and it builds the current sheet up; we did not succeed identify this in the present observations. 
 Then, the small amount of magnetic flux reconnected per unit of time, with its associated magnetic energy release, is expected to have a low contribution to the coronal emissions, as observed.

 We interpret the increase after $\approx$ 15:05 UT of the X-ray flux measured by GOES as an indication of significant magnetic reconnection, which accelerates energetic particles and then heats the plasma; this results in the evaporation of part of the chromospheric plasma, then an increase in the coronal density, and finally an enhancement to the EUV emission. This reconnection further decreases the stabilizing effect of the overlying magnetic field.  Then, a strong increase in the prominence speed (Fig. \ref{fig_timeslice_s1_s2}d) is a logical consequence since there is positive feedback on the dynamic of the reconnected field with a decrease in the downward magnetic tension \citep{Welsch2018}.
Further, a sudden transition in the kinematic evolution of the prominence from its slow to fast ascent together with the simultaneity in the increase of the HXR flux point toward the feedback process between the early dynamics of the eruption and the strength of the flare magnetic reconnection
\citep{Temmer2008, Vrsnak2016}.

The upward motion of coronal loops occurred before a significant upward motion of the prominence could be detected (at about 14:40 UT, middle and bottom panels of Figs. \ref{fig_timeslice_s1_s2} and \ref{fig_timeslice_s3_s4}).  
The upward motion of loops has been observed in other eruptive events and interpreted as the upward progression of the magnetic reconnection site into the corona due to the effect of the erupting magnetic structure, frequently with a filament present within, on the observed loops \citep{Liu2009b, Gosain2012, Simoes2013, Wang2018}. 
HXR coronal sources together with conjugate foot-point sources are a natural consequence of the coronal magnetic reconnection. In this process, the strongest sources are flare loop foot-point sources due to thick-target bremsstrahlung in or near the chromosphere
\citep{Aschwanden2002}.
As the erupting configuration goes on one side of the loops, they could be pushed sideways, then they finally retract as a consequence of a lower magnetic pressure in the region emptied by the erupting configuration.

For the present event, the earlier contraction of some coronal loops started at $\approx$ 14:59 UT (Fig. \ref{fig_timeslice_s3_s4}), so at least about 19 minutes after the start of the prominence eruption when the exponential amplitude was large enough to be detectable.
This result is in contrast to the expectation of \citet{Hudson2000};
since the eruption is not associated with any evidence of ``magnetic implosion'' for 19 minutes, then there is no evidence that the magnetic 
energy released is powered by such an implosion.
These observations are rather compatible with the theoretical models involving an ideal instability \citep[\eg\ ][and related models summarized in Sect. \ref{sect_Introduction}]{Lin2000}.
To put it simply, unstable magnetic configurations, for example,\ with an unstable FR, have a decreasing magnetic energy while they are in expansion. Furthermore, an instability can occur in models with an invariance along the PIL. These models  do not have any ``magnetic implosion'' while the FR is erupting and expanding.

The earlier contraction of some coronal loops is closer to the increase in the X-ray flux measured by GOES, and most of the observed contractions are in the time range of the X-ray flux enhancement observed by both GOES and RHESSI (15:10 -- 15:40 UT, Fig. \ref{fig_rhessi_lc}). This is in agreement with some earlier studies which found the loop contraction to be associated to the impulsive phase of some eruptive flares \citep{Liu2009,Simoes2013,Wang2018}.  In the present studied event, this time period also corresponds to the higher speed regime of the prominence (on the order of 200 \kms ).

Next, we notice a propagation of the contraction with height.
This result agrees with previous observations of other eruptions \citep[\eg\ ][]{Gosain2012,Simoes2013,Shen2014}.
In addition to an expected longer delay with an increasing distance to the launch site, a gradient of the Alfv\'en and fast mode speeds may be needed to explain the observations, as follows.  Both speeds have similar values in the low plasma $\beta$ of the corona.
A lower Alfv\'en speed at a lower height allows one to see the propagation of the contraction, while a much higher Alfv\'en speed at larger heights implies a propagation that is too fast in order to be detected with AIA time cadence (12 s).  Furthermore, this difference in speed could not only be due to a difference in height, but also to a difference in location in the AR (along the line of sight) of the short and large loops. Indeed, the projection along the line of sight still allows one to infer different loop shapes, and then different magnetic connectivities.
 
With a different view point, Fig. \ref{fig_timeslice_s3_s4} shows that, at the same time, loops expand or contract at different locations of the coronal configuration.  This is in agreement with previous results obtained for other eruptions \citep{Dudik2016, Dudik2017, Dudik2019}.
This evolution variety could be interpreted in the context of the vortex model derived from MHD simulations \citep{Zuccarello2017}.  The studied AR is suited for a comparison to these simulations because the large-scale distribution of the radial magnetic field component, at the photospheric level, is comparable to the set-up of the numerical simulations (i.e., bipolar with a moderately more disperse following magnetic polarity).  A first difference between the numerical simulations and observations is an expected magnetic Reynolds larger by orders of magnitude in the corona than in simulations, which is due to the limitations of current computer resources. A second difference is the earlier eruption observed on 01 March. This eruption opened, or at least made the loops large scale, in the northern side of the AR magnetic field (Fig. \ref{fig_evolution_171}).
The comparison of the present observations to the MHD simulations is limited by the few observed loops which are dense enough to be clearly visible, so their motion can be derived. Then, present observations do not allow us to visualize if a vortex is forming at the difference of MHD simulations where as many plasma blobs as needed could be followed in time. Then, we only claim that the studied loop evolution is compatible with the development of a lateral vortex, while more observations, in particular with a larger number of dense loops, are needed to confirm or refute this conclusion.

\section{Conclusions}
\label{sect_Conclusion}

We studied a prominence eruption which occurred at the solar western limb in AR  12290 on 02 March 2015. It is associated with an M3.7 GOES class flare. In the field of view of AIA, the initially stable prominence progressively accelerates, with an exponential behavior, to speeds in the range of 110 -- 240 \kms . This range corresponds to different plasma blobs and it traces the expansion of the magnetic configuration in the low corona. The prominence erupted away from the local vertical toward the northeast direction.  

 The source AR, which was observed during its disk transit, is a simple bipolar region in the decaying stage.
For about one week before the eruption, several episodes of magnetic flux cancellation were observed in between the polarities in the region where the filament and prominence was observed.
Sheared coronal loops were also observed with AIA. 
The AR magnetic configuration is asymmetric with a following polarity more extended than the leading one.
However, the coronal loop observations indicate that this does not create a significant asymmetry in the coronal connections. An earlier eruption created an open field- and high-lying loops configuration in the northern side of the AR about one day before the studied prominence eruption.
We conclude that this open- and high-lying field provides an easier channel for the prominence eruption, which implies a non-radial eruption.
 
AIA observations show that the prominence is embedded in a coronal structure which has the typical shape of an FR viewed along its local axis direction in its northern part. As the eruption develops, the FR-shaped emission increases in size and plasma following backward toward the chromosphere is observed at both prominence legs. 
Finally, RHESSI observations observed the foot-points and coronal sources of the flare loops.
We conclude that the event of 02 March 2015 has the main characteristics of an erupting FR.

The cold material emission of the prominence, observed in \ha, were split into two main blobs during the eruption.
The co-alignment with AIA observations show that one blob is at the rear side while the other one is at the front side of the erupting FR.
The blob at the rear side is at the expected location for dense plasma to be supported in the concave-up part of the FR.
The detection of the cold plasma in the front part of the FR is unusual.
This plasma is away of its equilibrium position in the magnetic dips.
Then, during the eruption, significant kinetic energy was inputed to drive it up to the top of the FR field lines.
Such cold plasma present in the front is found in 35\% of the cases of interplanetary
CMEs where cold plasma is detected in situ at 1 AU  \citep{Lepri2010}.
Then, our observations provide an alternative explanation to the one proposed by
\citet{Manchester2014}, where the strong deceleration of a fast CME by the interaction with the surrounding solar wind implies a drift of the dense and cold plasma toward the front of the CME.

The prominence eruption first initiates the expansion, and then the contraction of different sets of coronal loops located in the southern side of the eruption.  The expansion and contraction speeds of coronal loops are in the range of
2 \kms\ to 34 \kms.  Such results are coherent with the results obtained in other eruptions (Sect. \ref{sect_Introduction}). Here, with the analyzed eruption being at the limb, we take advantage of the fact that there is less background and foreground than with the previous analyzed eruptions observed on the solar disk.
This event also provides a different and useful view point directed almost along the FR axis in its northern part.

Our results show many similarities with previous studies of loop contraction in eruptions, see Sect. \ref{sect_Introduction}, and especially with the studies of \citet{Shen2014} and \citet{Dudik2016,Dudik2017}.
Similar to the latter studies, we observe a slow expansion of coronal loops before the onset of a filament and prominence eruption.  Significantly after the prominence eruption starts, loop contraction was detected on the southern side of the erupting FR.  This delay 
between prominence eruption and loop contraction onsets, of about 19 minutes, is well 
within the range of these three studies ($\approx$ 4, 7, 50 minutes). Also in agreement with \citet{Dudik2016,Dudik2017}, we simultaneously observe,  at different distances from the erupting FR,  sets of loops that are either in contraction or expansion.   
The expansion and contraction speed magnitudes are also comparable. 

The present studied eruption, as well as the observations cited in the previous paragraph could be interpreted as follows. First, due to the FR uplifting, the coronal loops were pushed upward. Afterwards, the loops began to contract in order to reach a new equilibrium state.
Another explanation is provided by the numerical simulations of \citet{Aulanier2010} and \citet{Zuccarello2017}.
In these simulations, the FR dynamics, linked to its instability, is the driver of the eruption.
Both in the present observations and  these simulations, the upward motion of
the FR starts before the loop contraction which occurs around the time when the FR strongly accelerates upward.  Hydrodynamic vortexes are generated which drive coronal loops upward or downward  depending on their spatial locations within the time-dependent vortexes.
A conclusion to further tests on a broader set of events where the dynamic of the magnetic configuration could be inferred from the beginning of the prominence and filament eruption, which is typically earlier than the associated flare. For that purpose, the presence of dense plasma in the largest possible fraction of the magnetic configuration is an important clue.

\begin{acknowledgements}
        We would like to thank the referee for the constructive comments and suggestions.
                We recognize the collaborative and open nature of knowledge creation and dissemination, under the control of the academic community as expressed by Camille No\^{u}s at http://www.cogitamus.fr/indexen.html.
SDO is a mission for NASA$'$s Living With a Star (LWS) Program. SDO data are courtesy of the NASA/SDO science team. 
RHESSI is a NASA Small Explorer Mission. PD acknowledges the support from CSIR, New Delhi.  
The work of RC is supported by the Bulgarian Science Fund under Indo-Bulgarian bilateral project.
RJ thanks the Department of Science and Technology (DST), Government of India for an INSPIRE fellowship and to CEFIPRA for a Raman Charpak Fellowship at Observatoire de Paris, Meudon, France.
BS would like to thank Ramesh Chandra for her stay in Nainital in October 2019 where this work was initiated.
\end{acknowledgements}

\bibliographystyle{aa}
\bibliography{reference}

\begin{appendix}
\section{Measured velocities}
\label{sect_Measured_Velocities}

The velocities are deduced from the time-distance analysis of the AIA observation shown in
Figs. \ref{fig_timeslice_s1_s2} and \ref{fig_timeslice_s3_s4}. These velocities are summarized in the two most right columns of Table \ref{tab_velocity}. They correspond to the loop systems "a" to "f" defined in Figs. \ref{fig_timeslice_s1_s2}b and \ref{fig_timeslice_s3_s4}b.  The mean velocities were computed from the slope found in the time-distance plot of panels (e) and (h) of Figs. \ref{fig_timeslice_s1_s2} and \ref{fig_timeslice_s3_s4} within the time range indicated in the third and fourth columns.

\begin{table*}[h!]
\centering
\caption{Expansion and contraction velocities of loops observed with AIA 304, 171, and 193 \AA\ filters.}
\label{tab_velocity}
\setlength{\tabcolsep}{18pt}

\begin{tabular}{c|c|c|c|c|c}
\hline
\centering
Loops & Wavelength & \multicolumn{2}{|c|}{Time (UT)} & \multicolumn{2}{c}{Velocity (km s$^{-1}$)} \\
\cline{3-6}
 & (\AA) & Expansion & Contraction & Expansion & Contraction \\
\hline
a  & 171 & 14:00 -- 15:12 & --           & 7 & -- \\
   &     & 14:00 -- 15:16 & --                   & 8 & --   \\
   &     & 14:00 -- 15:10 & --                   & 9 & --   \\
\cline{2-6}
   & 193 & 14:00 -- 15:16 & --           & 9 & -- \\
   &     & 14:22 -- 15:10 & --                   & 10& --   \\
\hline
b  & 304 & 14:08 -- 14:56 & 15:04 -- 15:42       & 8  & 11 \\
   &     & 14:36 -- 15:03 & 14:56 -- 15:12       & 9  & 14 \\
   &     & 14:28 -- 15:04 & 15:03 -- 15:34       & 11 & 19 \\
\cline{2-6}
   & 171 & --            & 15:10 -- 15:34       & -- & 9  \\
   &     & --            & 15:15 -- 15:37 &--  & 12 \\
   &     & --            & 15:03 -- 15:41 &--  & 12 \\
   &     & --            & 15:15 -- 15:34 & -- & 13 \\
   &     & --            & 15:04 -- 15:12 & -- & 18 \\
   &     & --            & 14:53 -- 15:15 & -- & 20 \\
   &     & --            & 14:55 -- 15:06 &-- & 22 \\
   &     & --            & 15:18 -- 15:37 &-- & 22 \\
   &     & --            & 15:00 -- 15:24 &-- & 22 \\
   \cline{2-6}
   & 193 &--             & 14:52 -- 15:24 &--    & 9 \\
   &     &--             & 15:11 -- 15:24 &--    & 9 \\
   &     &--             & 14:44 -- 15:27 & --   & 12 \\
   &     &--             & 15:18 -- 15:50 & --   & 21 \\
\hline
c   & 171 & --           & 15:14 -- 15:28        & -- & 6 \\
\cline{2-6}
   & 193 & --            & 15:18 -- 15:34        & -- & 6 \\
\hline
d   & 171 & 14:13 -- 15:12 & 15:18 -- 15:30      & 6  & 23 \\
\cline{2-6}
   & 193 & 14:20 -- 15:14       & 15:14 -- 15:28         & 6  & 17 \\
\hline
e  & 171 & 14:00 -- 15:08       & 15:13 -- 15:32         & 4  & 7 \\
   &     & 14:00 -- 15:06       & 15:12 -- 15:32         & 5  & 9 \\
   &     & --           & 15:14 -- 15:28         & -- & 23 \\
   &     & --           & 15:18 -- 15:38         & -- & 24 \\
   &     & --           & 15:18 -- 15:32         & -- & 31 \\
\cline{2-6}
   & 193 & 14:26 -- 15:06       & 15:20 -- 15:28         & 3  & 11 \\
   &     & 14:40 -- 15:08       & 15:20 -- 15:32   & 4  & 14 \\
   &     & 14:16 -- 15:08       & 15:27 -- 16:00         & 5  & 14 \\
   &     & 14:54 -- 15:16 & 15:14 -- 15:30   & 5  & 24 \\
   &     & --           & 15:14 -- 15:23         & -- & 30  \\
   &     & --           & 15:18 -- 15:30         & -- & 34  \\
\hline
f  & 171 & 14:00 -- 14:56       & --     & 2 & -- \\
   &     & 14:00 -- 15:10       & --     & 2 & -- \\
\cline{2-6}
   & 193 & 14:00 -- 16:20       & --     & 2 & -- \\
\hline
\end{tabular}
\end{table*}

\end{appendix}

\end{document}